\documentclass[%
reprint,  
 amsmath,amssymb,
 aps,
]{revtex4-2}
\usepackage{enumitem}
\usepackage{graphicx}% Include figure files
\usepackage{dcolumn}% Align table columns on decimal point
\usepackage{bm}% bold math
\usepackage{hyperref}% add hypertext capabilities
\usepackage{slashed}
\usepackage{subfloat}
\usepackage{multirow}
\usepackage[table]{xcolor}
\usepackage{pgfplots,subfigure}
\usepackage{orcidlink}
\usepackage{float}
\usepackage{geometry}
\usepackage[table]{xcolor} % Add in preamble
\usepackage{booktabs}       % Add in preamble
\usepackage{array}          % For custom column types
\usepackage{hyperref}
\usepgfplotslibrary{fillbetween}

\newcolumntype{C}{>{\centering\arraybackslash}m{4cm}} 
\definecolor{acsblue}{RGB}{17,76,139}

\begin{document}

\fontsize{7.6}{8.6}\selectfont
%\preprint{APS/123-QED}
\title{Schr\"odinger and Klein-Gordon oscillators in Eddington-inspired Born-Infeld gravity: Degree-one Confluent Heun polynomial  correspondence}

\author{Omar Mustafa\orcidlink{0000-0001-6664-3859}}
\email{omar.mustafa@emu.edu.tr }
\affiliation{Department of Physics, Eastern Mediterranean University, 99628, G. Magusa, north Cyprus, Mersin 10 - Türkiye}

\author{Abdullah Guvendi\orcidlink{0000-0003-0564-9899}}
\email{abdullah.guvendi@erzurum.edu.tr}
\affiliation{Department of Basic Sciences, Erzurum Technical University, 25050, Erzurum, Türkiye}

\date{\today}

\begin{abstract}
{\fontsize{7.6}{8.6}\selectfont \setlength{\parindent}{0pt}  
We investigate Schr\"odinger and Klein-Gordon (KG) oscillators in the spacetime of a global monopole (GM) within Eddington inspired Born-Infeld (EiBI) gravity, including, in the relativistic sector, the coupling to a Wu-Yang magnetic monopole (WYMM). By reducing the radial equations to the confluent Heun form and enforcing termination of the Heun series, we obtain conditionally exact solutions in which the radial eigenfunctions truncate to polynomials of degree $(n+1)\geq 1$. This truncation imposes algebraic constraints \textcolor{red}{on} the oscillator frequency and restricts the \textcolor{red}{allowed maximum value for the} orbital angular \textcolor{red}{momentum quantum number} $\ell$ \textcolor{red}{for each $n\geq 0$}. In the lowest nontrivial case $n=0$, the degree-one Heun polynomial yields a closed analytic expression for the frequency and determines a finite upper bound on $\ell$, dictated jointly by the EiBI deformation and the GM deficit. The resulting parametric correlations reveal a sharp geometric control of the spectrum: EiBI nonlinearities and the angular deficit fix the admissible bound states through polynomial truncation conditions. The confluent Heun correspondence is made explicit, providing a rigorous and reproducible framework for extracting analytical solutions from otherwise non-polynomial Heun structures. Applying the same method to the KG oscillator with a WYMM, we derive conditionally exact particle and antiparticle energies in a closed form. The relativistic spectrum exhibits perfect charge symmetry and a precise dependence on the WYMM strength, the EiBI parameter and the angular momentum constraint. To the best of our knowledge, this constitutes the first unified and fully consistent treatment of conditionally exact Schr\"odinger and Klein-Gordon \textcolor{red}{oscillator solutions} in EiBI gravity based on a degree-one confluent Heun polynomial.
}
\end{abstract}


\maketitle

\tableofcontents

\setlength{\parindent}{0pt}

\section{Introduction}\label{sec:1}

\setlength{\parindent}{0pt}

Gravity is inherently portrayed as a geometric manifestation of spacetime by Einstein's pioneering groundbreaking theory of general relativity (GR) \cite{rf1}. GR has sparked significant interest in profound investigations on the intricate interplay between gravitational fields and quantum mechanical phenomena, captivating in turn eloquent correlation between spacetime-induced gravitational force fields and quantum mechanical spectroscopic structures of particles, offering deeper insights into the understanding of the fundamental nature of the primordial universe. Due to the phase transition of the early universe, topological defects are formed \cite{rf2,rf3}, encapsulating domain walls, cosmic strings, and global monopoles \cite{rf4,rf5,rf6,rf7,rf8,rf9}.

\setlength{\parindent}{0pt}

The theory of Eddington-inspired Born-Infeld (EiBI) gravity, on the other hand, integrates Born-Infeld nonlinear electrodynamics with Eddington gravitational action, modifying in effect general relativity (GR) \cite{int1,int2,int3,int4}. While Einstein's GR in vacuum is a natural reduction of EiBI gravity, the latter maintains internal consistency and avoids instabilities and ghosts \cite{int5}. EiBI prevents, moreover, cosmological singularities and allows a singularity-free cosmological evolution that facilitates the existence of compact objects like neutron stars and other stellar structures that have been subject to different studies for nuclear astrophysics \cite{int6,int7,int7a,int8,int9,int10,int11}. Nevertheless, it offers alternative gravitational explanations in the strong field and high energy regimes, allowing asymptotic compatibility with GR even at the classical level \cite{int12,int1,int4,int11,int5}.

\setlength{\parindent}{0pt}

The metric describing a global monopole (GM) spacetime in EiBI is given by
\begin{equation}
ds^{2}=-\tilde{\alpha} ^{2}\,dt^{2}+\frac{r^{2}}{\tilde \alpha ^{2}\left( r^{2}+\kappa
\tilde{\beta} \right) }\,dr^{2}+r^{2}\,\left( d\theta ^{2}+\sin
^{2}\theta \,d\varphi ^{2}\right) ,  \label{2.1}
\end{equation}
where $\kappa$ is the Eddington parameter that controls nonlinearity, $0<\tilde{\alpha}^{2}=1-\tilde{\beta} \leq 1$, with $\tilde{\beta} =8\pi G \eta^{2}$ being the deficit angle, $\tilde\alpha$ the GM parameter depending on the energy scale $\eta$ and $G$ Newton’s constant \cite{rf10,rf11,rf12,rf13,rf14,rf15,rf16,rf17,rf18,rf19,rf20,rf21,rf22,rf23,rf24,rf25,rf26}. Upon rescaling of $\sqrt{\left( 1-\tilde\beta \right) }\,dt\rightarrow dt$, along with substituting $\tilde{\kappa}=\kappa \tilde\beta $, one may recast (\ref{2.1}) as 
\begin{equation}
ds^{2}=-\,dt^{2}+\frac{1}{\tilde\alpha ^{2}\left( 1+\frac{\tilde{\kappa}}{r^{2}}
\right) }\,dr^{2}+r^{2}\,d\theta ^{2}+r^{2}\sin ^{2}\theta \,d\varphi ^{2}.
\label{2.2}
\end{equation}
It could be interesting to know that $\tilde{\kappa}<0$ describes a topologically charged wormhole \cite{rf27,rf28,rf29}, $\tilde\alpha =1$ and $\tilde{\kappa}<0$ correspond to a Morris-Thorne-type wormhole spacetime \cite{rf30,rf31}, $\kappa =0$ describes a GM spacetime without EiBI gravity, and $\tilde\kappa >0$ corresponds to a GM spacetime in EiBI gravity. GM spacetime in EiBI gravity is the background of our interest in the current study. This metric has non-vanishing elements given by \[g_{tt}=-1,\,\,g_{rr} =\frac{r^{2}}{\tilde\alpha ^{2}\left( r^{2}+\tilde{\kappa}\right) }
,\,\,g_{\theta\theta}=r^{2},\,\,g_{\varphi \varphi }=r^{2}\sin ^{2}\theta,\]so that their inverses are given by
\begin{equation}
g^{tt}=-1,\quad g^{rr} =\frac{\tilde\alpha ^{2}\left( r^{2}+\tilde{\kappa}\right) }{r^{2}}
,\quad g^{\theta\theta}=\frac{1}{r^{2}},\quad g^{\varphi \varphi }=\frac{1}{r^{2}\sin ^{2}\theta }.
\label{2.3}
\end{equation}
On the other hand, intricate and intriguing effects of different gravitational force fields generated by spacetime backgrounds on the spectroscopic structure of quantum particles (both relativistic and non-relativistic) have attracted research attention over the years. In particular, harmonic oscillators in different spacetime backgrounds form a fundamental model study of interest for quantum gravity, and  geometric theory of topological defects in condensed-matter physics (e.g., \cite{Ref29,Ref30,Ref31,Ref32}). It could therefore be interesting to recollect a few critical points. 

We start with introducing the non-relativistic Schr\"odinger  harmonic oscillator in different spacetime backgrounds through the non-minimal coupling $\partial_j\to \partial_j \pm \mathcal{F}_j$, where $\mathcal{F}_j \equiv \left( \mathcal{F}_r, 0, 0 \right)$ and $\mathcal{F}_r = m \omega r$. In this case, the Schr\"{o}dinger oscillator, in $\hbar=2m=1$ units, reads
\begin{equation}
\left[ -\left( \frac{1}{\sqrt{g}}(\partial_i + \mathcal{F}_i)\sqrt{g}%
\,g^{ij}\,(\partial_j - \mathcal{F}_j)\right) -E\right] \,\Psi \left(r,\theta ,\varphi \right) =0,  \label{2.5}
\end{equation}
where, in such units, $\mathcal{F}_r=\omega r/2$, $i=r,\theta, \varphi$, and
\begin{equation}
g=\det \left( g_{ij}\right) =\frac{r^{6}\sin ^{2}\theta }{\tilde\alpha ^{2}\left(
r^{2}+\tilde{\kappa}\right) }\Rightarrow \sqrt{g}=\frac{r^{3}\sin \theta }{%
\tilde\alpha \sqrt{r^{2}+\tilde{\kappa}}}.  \label{2.4}
\end{equation}
Therefore, it is obvious that the curved spacetime background modifies the form of  Schr\"{o}dinger oscillator. At this point, one should observe that such a non-minimal coupling is manifestly inherited from the textbook Schr\"odinger oscillator annihilation and creation operators \cite{Fluge}. The same recipe is applied to the Dirac and KG oscillators \cite{rf32,rf33} so the KG-oscillator equation, with $c=1=\hbar$, is given by
\begin{equation}
    \frac{1}{\sqrt{-g}} (\partial_\mu +\mathcal F_\mu)\sqrt{-g}g^{\mu\nu}(\partial_\nu -\mathcal F_\nu)\Psi \left(t,r,\theta ,\varphi \right)=m_\circ^2 \Psi \left(t,r,\theta ,\varphi \right), \label{1.1}
\end{equation}
where $\mu,\nu=t,r,\theta, \varphi,\,\mathcal F_\mu=(0,\mathcal F_r,0,0)$ and 
\begin{equation}
g=\det \left( g_{\mu\nu}\right) =-\frac{r^{6}\sin ^{2}\theta }{\tilde\alpha ^{2}\left(
r^{2}+\tilde{\kappa}\right) }\Rightarrow \sqrt{-g}=\frac{r^{3}\sin \theta }{%
\tilde\alpha \sqrt{r^{2}+\tilde{\kappa}}}.  \label{1.2}
\end{equation}
Hereby, one should be very reluctant to use $\mathcal F_r=m_\circ \omega r$ since $m_\circ=m_\circ c^2$ (rest mass energy that yields dimensionally inconsistent KG-oscillator equation) in (\ref{1.1}) but should rather use $\mathcal F_r=\Omega r$ to produce a KG-oscillator (c.f.,e.g.,  \cite{rf25,rf26} and section \ref{sec:3} below). It is obvious that this recipe incorporates the effect of different spacetime fabrics in the structures of the Schr\"odinger (\ref{2.5}) and KG (\ref{1.1}) oscillators.

\setlength{\parindent}{0pt}

In the current methodical proposal, we discuss the Schr\"odinger (in section \ref{sec:2}) and KG (in section \ref{sec:4}) oscillators in an EiBI gravity background. In the KG oscillator case, we also \textcolor{red}{include a Wu-Yang magnetic monopole (WYMM)} \cite{rf21,rf26,rf331,rf332}. For the Schr\"odinger oscillator, in section \ref{sec:2}, we discuss a power series solution and truncate it to a polynomial of degree $(n+1) \geq 1$ to obtain a conditionally exact solution \textcolor{red}{(i.e., analytical solutions exist only when some or all effective potential parameters are fine-tuned to satisfy specific constraints or correlations \cite{rf333})}. \textcolor{red}{In the same section, we also analyze} a non-trivial case with $n=0$ that represents a polynomial of degree one solution. 
\textcolor{red}{Within this framework, we show that not only the oscillator frequency is explicitly $n$-dependent (metaphorically quantized, in the sense that each $n$-state corresponds to a distinct oscillator frequency), but also the angular momentum quantum number $\ell$ is restricted to a maximum allowed value, dictated by the parametric correlations.}
In section \ref{sec:3}, we report the confluent Heun series/polynomial correspondence and report the conditional exact solvability for a degree one confluent Heun polynomial. We believe that this correspondence is necessary and useful for both the readers and researchers  who are only interested in the non-trivial confluent Heun polynomial solution of degree one, a common practice/procedure in the recent literature. In section \ref{sec:4}, we follow the conditionally exact  confluent Heun polynomial of degree one solution and report the KG oscillator particle and antiparticle energies. To the best of our knowledge, such a conditionally exact solution procedure has never been reported elsewhere in the literature. Our concluding remarks are given in section \ref{sec:5}.

\section{Schr\"odinger oscillator in E\MakeLowercase{i}BI gravity spacetime}\label{sec:2}

\setlength{\parindent}{0pt}
 
We now consider the Schr\"{o}dinger oscillator, $\mathcal{F}_r=\omega r/2$, in EiBI-gravity so that Eq. (\ref{2.5}) would then imply
\begin{equation}
    \begin{split}
        &R''+\left(\frac{1}{r}+\frac{r}{r^2+\tilde{\kappa}}\right)R'+\left[\frac{\mathcal{E}r^2}{r^2+\tilde{\kappa}}-\frac{1}{4}\omega^2 \,r^2-\frac{\mathcal{L}(\mathcal{L}+1)}{r^2+\tilde{\kappa}}-\omega\right]R=0, \label{2.6}
    \end{split}
\end{equation}
where $\mathcal{E}=E/\tilde\alpha^2-\omega/2$, and
\begin{equation}
    \mathcal{L}(\mathcal{L}+1)=\ell(\ell+1)/\tilde\alpha^2\Rightarrow \mathcal{L}=-\frac{1}{2}+\frac{1}{2\tilde\alpha}\sqrt{4\ell^2+4\ell+\tilde\alpha^2}. \label{2.7}
\end{equation}
Let us now use the substitution
\begin{equation}
    R(r)= \, \exp(
    {-\frac{\omega r^2}{4}})\,H(r), \label{2.8}
\end{equation}
followed by the change of variable $s=r^2/\tilde k+1\Rightarrow r=\sqrt{\tilde k (s-1)}$ to obtain
\begin{equation}
\left(s-1 \right) H''(s)+\left[-\frac{\omega \tilde k}{2}s+P-\frac{1/2}{s}\right]H'(s)
      +\left[(\mu+\nu)-\frac{\mu}{s}\right]H(s)=0,
 \label{2.9}
  \end{equation}
where 
  \begin{equation}
      \begin{split}
          P&=\frac{\omega \tilde k}{2}+\frac{3}{2},\,\, \tau=\mathcal L (\mathcal L+1), \,\,\tilde E=\frac{E}{\tilde \alpha^2},\,\,\nu=-\frac{1}{4}\left(\tau+2\tilde \kappa \omega\right)\\ \\
\mu&=\frac{1}{4}\left[\tilde E \tilde \kappa-\tilde \kappa \omega+\tau\right],\,\,\mu+\nu=\frac{\tilde\kappa}{4}\left(\tilde E  -3\omega\right)
      \end{split} \label{2.9.1}
  \end{equation}
Next, we use a power series expansion in the form of
  \begin{equation}
      H \left( s\right) =\sum_{j=0}^{\infty }A_{j}\,s^{j}.\label{2.10}
  \end{equation}
Hereby, it should be noted that \[R(r)=\frac{U(r)}{r}=
exp(-\frac{\omega r^2}{4})\,H(r)\,\Rightarrow \lim\limits_{r\rightarrow 0}\,U(r)\rightarrow 0 ,\]is a condition that should be enforced on the radial wave function $U(r)$, whilst $R(r)$ should at least be finite as $r\to 0$. Moreover, using (\ref{2.10}) in (\ref{2.9}), one would also obtain
\begin{equation}
A_1=-2\mu A_0;\quad A_0=1, \label{2.10.1}
\end{equation}
along with a three term recursion relation in the form of
\begin{equation}
\begin{split}
A_{j+2}\, \left( j+2\right)\left( j+\frac{3}{2}\right) &=A_{j+1}\left[ \left(
j+1\right) \left( j+P\right)-\mu \right]\\
&+A_{j}\left[ \mu+\nu-\frac{\tilde \kappa \omega}{2}j\right]=0. \label{2.11} 
\end{split}
\end{equation}
Where
\begin{equation}
    \begin{split}
        &A_2 (3)=A_1[P-\mu]+A_0[\mu+\nu]\\
        &A_3\left(15/2\right)=A_2 [2(P+1)-\mu]+A_1\left[\mu+\nu-\frac{\tilde \kappa\omega}{2}\right],\\
        &\vdots.
    \end{split}\label{2.11.1}
\end{equation}
Now, we wish to truncate the power series to a polynomial of degree $(n+1)\geq1$ to secure finiteness and square integrability of the wave function. In this case, we need  the $(n+1)^{th}$ order recursion relation, that is, for $j=(n+1)$ the three term recursion relation reads
\begin{equation}
\begin{split}
A_{n+3}\, \left( n+3\right)\left( n+\frac{5}{2}\right) &=A_{n+2}\left[ \left(
n+2\right) \left( n+P+1\right)-\mu \right] \\ 
&+A_{n+1}\left[ \mu+\nu-\frac{\tilde \kappa\omega}{2} (n+1)\right]=0;\\
&\,\, n=0,1,2,\cdots.
\end{split} \label{2.12}
\end{equation}
The truncation of the power series to a polynomial of degree $(n+1)$ is achieved by enforcing the conditions that $A_{n+2}=0=A_{n+3}$ and $A_{n+1}\neq 0$. This would imply that
\begin{equation}
 \mu+\nu=\frac{\tilde \kappa w}{2}(n+1)\Rightarrow =\frac{\tilde\kappa}{4}\left(\tilde E  -3\omega\right)=\frac{\tilde \kappa w}{2}(n+1);\quad 0\leq n=0,1,2,\cdots,   \label{2.12.1}
\end{equation}
to imply \[E=2\omega \tilde \alpha^2 \left(n+\frac{5}{2}\right)\Rightarrow E_n=2\omega \tilde \alpha^2 \left(n+\frac{5}{2}\right).\]Under such truncation conditions, our power series truncates to a polynomial of degree $(n+1)$ and reads
\begin{equation}
      H \left( s\right) =\sum_{j=0}^{n+1 }A_{j}\,s^{j}\,;\,\,0\leq j \leq (n+1).\label{2.12.2}
  \end{equation}
However, in the process, the $n^{th}$ order recursion relation for  $j=n$, with $A_{n+2}=0$, reads
\begin{equation}
\begin{split}
0 &=A_{n+1}\left[ \left(n+1\right) \left( n+P\right)-\mu \right] 
+A_{n}\left[ \mu+\nu-\frac{\tilde \kappa\omega}{2} n\right]=0,\\ 
\Rightarrow & \quad A_{n+1}=-\frac{(\tilde \kappa\omega/2)}{(n+1)(n+3/2)+\nu}A_n,
\end{split}\label{2.13}
\end{equation}
and identifies yet another critical recursion relation that facilitates conditional exact solvability of the problem at hand. One should also be aware that the parametric relation \( A_1=-2\mu A_0\) in (\ref{2.10.1}) holds for all values of $n$.  This is to be illustrated in the subsequent non-trivial example.

\subsection{The non-trivial lowest state with $n=0\Rightarrow j=0,1$}\label{sec:2:1}

\setlength{\parindent}{0pt}

For the non-trivial lowest state $n=0$ and hence $j=0,1$, we have from (\ref{2.13}), together with \(   A_1=-2\mu_0 A_0\) in (\ref{2.10.1}), \[A_1=-\frac{(\tilde \kappa\omega/2)}{3/2+\nu}A_0\Rightarrow 2\mu_0 =\frac{ \tilde \kappa\omega}{3+2\nu}\]
by (\ref{2.12.1}) and (\ref{2.13}), noting that \(   A_1=-2\mu_n A_0;\, \forall n\geq0\). This in turn would imply a correlation between the oscillator frequency $\omega_0$, the angular momentum quantum number $\ell$, the GM parameter $\tilde \alpha$, and the Eddington parameter $\tilde\kappa$ given by
\begin{equation}
   \omega_{_0}=\frac{1}{8\tilde \kappa}\left[10-5 \tau+\sqrt{ \tau^2-12 \tau+100}  \right]\geq0; \,\, \tau=\frac{\ell(\ell+1)}{\tilde \alpha^2}\geq0. \label{2.14}
\end{equation}
This, in effect, identifies the conditionally exact solvability of the problem. Therefore, the lowest energy state, with $n=0$, is given by
\begin{equation}
     E_0=5\omega_{_0} \tilde \alpha^2=\frac{5\,\tilde \alpha^2 }{8\tilde \kappa}\left[10-5 \tau+\sqrt{ \tau^2-12 \tau+100}  \right]. \label{2.15}
\end{equation}

\begin{figure*}[ht!]  
\centering
\includegraphics[width=0.85\textwidth]{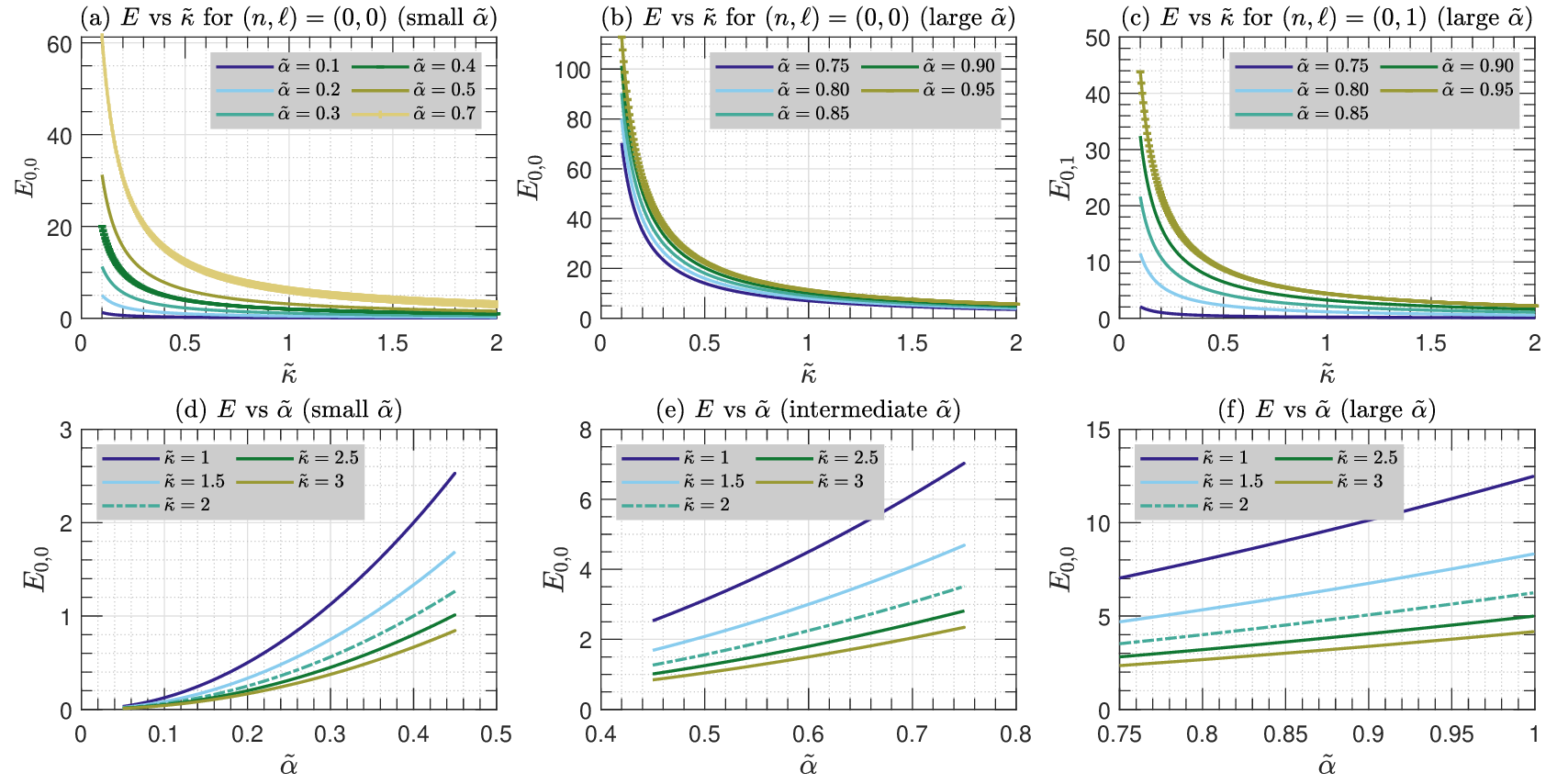}
\caption{\fontsize{7.6}{8.6}\selectfont $n=0$ energy states $E_{0,\ell}$ for Schr\"odinger oscillator in EiBI GM spacetime, computed from Eq.~(\ref{2.15}). Panels (a)--(c) show the dependence on the gravitational parameter $\tilde{\kappa}$ for fixed effective coupling $\tilde{\alpha}$; panels (d)--(f) show the dependence on $\tilde{\alpha}$ for fixed $\tilde{\kappa}$.}
\label{fig:1}
\end{figure*}
Notably, the result in (\ref{2.14}) is a parametric correlation that puts delicate restrictions on the allowed values for the angular momentum quantum number $\ell$, mandated by the allowed values of the GM parameter $0<\tilde{\alpha}^{2}\leq 1$, the Eddington parameter $\tilde \kappa$ and the oscillator frequency  $\omega_\circ> 0$. That is,
\begin{equation}
\begin{split}
   & \left[10-5 \tau +\sqrt{ \tau^2-12 \tau+100}  \right]\geq0\Rightarrow \tau_{max}<11/3\\ \\
   & \Rightarrow \quad \ell_{max}=-\frac{1}{2}+\sqrt{\frac{1}{4}+\frac{11}{3}\tilde\alpha^2}.
    \end{split} \label{2.16}
\end{equation}
On the other hand, since the allowed values of the GM parameter are given as \(0<\tilde\alpha^2\leq1\), one would find that $\ell_{max}\sim 1.479$, for $\tilde\alpha=1$. Hence, the maximum value for the angular momentum quantum number is an integer and reads $\ell_{max}=1$. Strictly speaking, for \(0< \tilde\alpha<\sqrt{66}/11\) the only allowed value is $\ell=0$ and for \(\sqrt{66}/11\leq \tilde\alpha<1\) the allowed values are $\ell=0,1$.

\setlength{\parindent}{0pt}

In Figure \ref{fig:1}, we meticulously follow the parametric restrictions mentioned above and show the Schr\"odinger oscillator energies. In Figures \ref{fig:1}(a), \ref{fig:1}(b), and \ref{fig:1}(c) we show the oscillator energies against the EiBI gravity parameter $\tilde\kappa$ for different GM parameter $\tilde\alpha$, where \ref{fig:1}(a) is plotted for  $0<\tilde\alpha<\sqrt{66}/11\simeq 0.738$ which only allows $\ell=0$, and \ref{fig:1}(b) and \ref{fig:1}(c) are plotted for $0.74<\tilde\alpha<1$ which allows $\ell=0$ and $\ell=1$, respectively. In figures \ref{fig:1}(d), \ref{fig:1}(e), and \ref{fig:1}(f) we show the oscillator energies against the GM parameter $\tilde\alpha$ for different values of the EiBI parameter $\tilde\kappa$, where \ref{fig:1}(d) is plotted for $0<\tilde\alpha<0.738$ that only allows $\ell=0$, and \ref{fig:1}(e) and \ref{fig:1}(f) are for $0.74<\tilde\alpha<1$ that allows $\ell=0$ and $\ell=1$, respectively.  The figures reflect exact behaviours as suggested by (\ref{2.15}) so that energy increases with increasing $\tilde\alpha$ and decreases with increasing $\tilde\kappa$.

\section{Confluent Heun polynomials of degree $(n+1)$ correspondence}\label{sec:3}

\setlength{\parindent}{0pt}

Here, we provide the corresponding confluent Heun series/polynomials. In so doing, we recall Eq.(\ref{2.9}) and rewrite it as
\begin{equation}
\left(s-1 \right) H''(s)+\left[\alpha s+P-\frac{(\beta+1)}{s}\right]H'(s)
      +\left[(\mu+\nu)-\frac{\mu}{s}\right]H(s)=0,
 \label{2.16}
  \end{equation}
where $\alpha =-\tilde \kappa \omega/2,\, \beta=-1/2$, 
  \begin{equation}
      \begin{split}
          P&=\beta+\gamma+2-\alpha=\tilde \kappa \omega/2+3/2\Rightarrow \gamma=0,\\\\
          \mu&=\frac{1}{2}\left[\alpha-\beta-\gamma+\beta(\alpha-\gamma)\right]-\eta=\frac{1}{4}\left[\tilde E \tilde \kappa-\tilde \kappa \omega+\tau\right], \\ \\
          \nu&=\frac{1}{2}\left[\alpha+\beta+\gamma+\gamma(\alpha+\beta)\right]+\delta+\eta=-\frac{1}{4}\left(\tau+2\tilde \kappa \omega\right).
      \end{split}\label{2.17}
  \end{equation}
In this case, Eq. (\ref{2.16}) admits a solution in the form of confluent Heun polynomials, so that \[H(s)=H_{C}\left(\alpha,\,\beta,\,\gamma,\,\delta,\,\eta, s\right)=\sum_{j=0}^{n+1 }A_{j}\,s^{j}\,;\,\,0\leq j \leq (n+1).\]Consequently, the $(n+1)^{th}$ order recursion relation reads
  \begin{equation}
  \begin{split}
   &\mu+\nu=\frac{\tilde \kappa\omega}{2}(n+1)=-\alpha(n+1) \\
   \Rightarrow \, & \delta=\delta_n=-\alpha\left[n+1+\frac{1}{2}\left(\beta+\gamma+2\right)\right];\, n=0,1,2,\cdots. 
   \end{split}\label{2.18}
  \end{equation}
At this point, it should be observed that $\delta$ is in perfect accordance with that reported in Ronveaux \cite{rf34} as \[\delta=-\alpha\left[\tilde n+\frac{1}{2}\left(\beta+\gamma+2\right)\right],\]where $\tilde n=n+1\geq1$ is a positive integer. Moreover, the $n^{th}$ order recursion relation \[A_{n+1}\left[ \left(n+1\right) \left( n+P\right)-\mu \right] 
+A_{n}\left[ \mu+\nu+\alpha n\right]=0,\]along with \[A_1=-\frac{\mu}{\beta+1} A_0;\quad A_0=1,\]determine a corresponding parametric correlation similar to that in (\ref{2.13}).

\subsubsection{Degree-one Confluent Heun polynomial correspondence}\label{sec:3:1}

\setlength{\parindent}{0pt}

Next, for the case $n=0$ we have our confluent Heun polynomial of degree one in the form of \[H_0(s)=H_{C,_0}\left(\alpha,\,\beta,\,\gamma,\,\delta,\,\eta, s\right)=A_0+A_1 s.\]A straightforward textbook substitution of $H_0(s)$ in (\ref{2.16}) would yield
\begin{equation}
\begin{split}
&\text{for $s^2$ coefficient}\Rightarrow \mu_0+\nu=-\alpha \Rightarrow \mu_0+\nu=\frac{\tilde \kappa\omega}{2} \\
&\text{for $s^1$ coefficient}\Rightarrow(P-\mu_0)A_1=-(\mu_0+\nu)A_0 \Rightarrow A_1=-\frac{\tilde \kappa\omega\textcolor{red}{/2}}{P-\mu_0}A_0 \\
&\text{for $s^0$ coefficient}\Rightarrow A_1(\beta+1)=-\mu_0 A_0 \Rightarrow A_1=-2\mu_0 A_0.
\end{split} \label{2.19}
\end{equation}
The combination of the second and third lines in (\ref{2.19}) would give the parametric correlation in (\ref{2.14}) and the first line yields \[ E_0=5\omega_0 \tilde \alpha^2=\frac{5\,\tilde \alpha^2 }{8\tilde \kappa}\left[10-5 \tau+\sqrt{ \tau^2-12 \tau+100}  \right].\]This result is in exact agreement with that reported in (\ref{2.14}) and (\ref{2.15}). This in turn would emphasis the restriction on the angular momentum quantum number discussed in (\ref{2.16}).

\section{KG-oscillators in E\MakeLowercase{i}BI gravity spacetime }\label{sec:4}

\setlength{\parindent}{0pt}

In this section, we closely follow Mustafa et al. \cite{rf26} who have considered KG-oscillators in EiBI gravity and a WYMM. We therefore skip their derivations and recollect their result reported in Eq. (22) which reads, with $\mathcal F_r=\Omega r$,
\begin{equation}
(r^2+\tilde \kappa) H''(r)+\left[2(1-\Omega \tilde\kappa)r+\frac{\tilde\kappa}{r}-2\Omega r^3\right]H'(r)+\left[P_1r^2-P_2\right]H(r)=0, \label{3.1}    
\end{equation}
where, \[\mathcal E^2=\frac{E^2-m_\circ^2}{\tilde\alpha^2}-3\Omega-\Omega^2\tilde\kappa, \quad \tilde\ell(\tilde\ell+1)=\frac{\ell(\ell+1)-\sigma^2}{\tilde\alpha^2}+2\Omega\tilde\kappa\]\[P_1=\mathcal E^2+\Omega^2\tilde\kappa-3\Omega,\,\,P_2=\tilde\ell(\tilde\ell+1)+2\Omega\tilde\kappa.\]
with $\sigma=eg$, $g$ is the WYMM strength. For more comprehensive details on such derivations, the reader may refer to Mustafa et al. \cite{rf26}. Next, we use a change of variables in the form of $s=r^2/\tilde\kappa-1$, in (\ref{3.1}), to obtain the confluent Heun equation in (\ref{2.16})  with the corresponding parameters
\begin{equation}
    \begin{split}
        &\alpha=-\Omega \tilde\kappa,\quad \beta=-\frac{1}{2},\quad P=\Omega \tilde\kappa+\frac{3}{2},\\
        &\mu+\nu=\frac{P_1\tilde\kappa}{4}, \quad\mu=\frac{P_1\tilde\kappa}{4}+\frac{P_2}{4} \Rightarrow\nu=-\frac{P_2}{4}.
    \end{split} \label{3.2}
\end{equation}

\begin{figure*}[ht!]  
\centering
\includegraphics[width=1\textwidth]{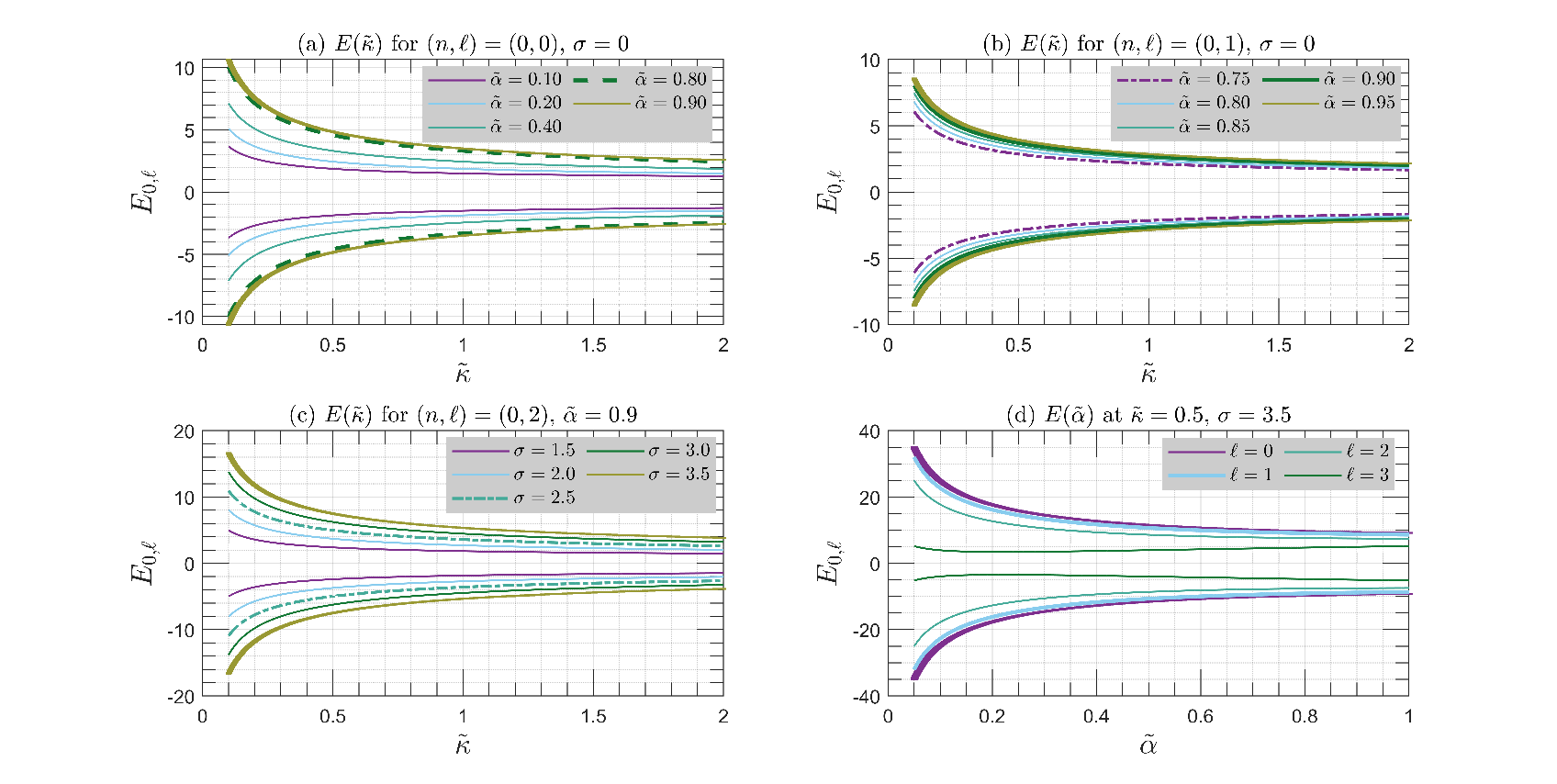}
\caption{\fontsize{7.6}{8.6}\selectfont  Energy levels of the KG oscillator in EiBI gravity with the WYMM extension.  
Each panel displays both positive and negative relativistic branches of the energy eigenvalues obtained from the analytic expression \eqref{3.5}. (a) Energy levels $E(\tilde{\kappa})$ for the ground state $(n,\ell)=(0,0)$ at $\sigma=0$ and several values of the deformation parameter $\tilde{\alpha}$. (b) $E(\tilde{\kappa})$ for $(n,\ell)=(0,1)$ at $\sigma=0$ and different $\tilde{\alpha}$ values, showing the influence of the angular momentum barrier. (c) $E(\tilde{\kappa})$ for $(n,\ell)=(0,2)$ with fixed $\tilde{\alpha}=0.9$ and increasing WYMM parameter $\sigma$, showing strong sensitivity of the spectrum to $\sigma$. (d) Energy levels as functions of $\tilde{\alpha}$ at fixed $\tilde{\kappa}=0.5$ and $\sigma=3.5$ for $\ell=0,1,2,3$.}
\label{fig:2}
\end{figure*}%
Hence, the substitution of the confluent Heun polynomial of degree one,  \(H_{C,_0}(s)=H(s)=A_0+A_1 s\), would yield
\begin{equation}
    \begin{split}
        &\mu_0+\nu=\frac{P_1\tilde\kappa}{4} \Rightarrow E_0=\pm\sqrt{10\,\tilde\alpha \,\Omega+m_\circ^2}\\
       & \mu_0+\nu=-\alpha \Rightarrow \mu_0=\frac{3}{2}\Omega \tilde\kappa+\frac{\tilde\ell(\tilde\ell+1)}{4}\geq0.\\
       &  \frac{\alpha}{P-\mu_0}=-\frac{\mu_0}{(\beta+1)} \Rightarrow  \mu_0=\frac{P}{2}+\frac{1}{2}\sqrt{\left(\Omega \tilde\kappa+\frac{1}{2}\right)^2+4}. 
    \end{split} \label{3.3}
\end{equation}
It is obvious that the first line of (\ref{3.3}) gives the energies for KG oscillators (particles and antiparticles), while the second and third lines should be combined to obtain the parametric correlation for \(\Omega=\Omega_0 \geq0\):
\begin{equation}
\Omega_0={\frac {3{\sigma}^{2}+10{\tilde \alpha }^{2}-3\tilde \tau+\sqrt 
{100{\tilde \alpha }^{4}+ \left( 12{\sigma}^{2}-12\tilde\tau \right) {
\tilde \alpha }^{2}+ \left( {\sigma}^{2}-\tilde \tau \right) ^{2}}}{16{\tilde \alpha }^{2}
\tilde\kappa}},\label{3.4}
\end{equation}
where $\tilde\tau=\ell(\ell+1)$, and hence
\begin{equation}
    E_0=\pm\sqrt{10\,\tilde\alpha \,\Omega_0+m_\circ^2}, \label{3.5}
\end{equation}
represents the energies of particles and antiparticles for the KG-oscillator in EiBI gravity and a WYMM. Obviously, this result offers a parametric correlation between the oscillator frequency $\Omega_0$, the Eddington parameter $\tilde \kappa$, the GM parameter $\tilde\alpha$, the angular momentum quantum number $\ell$, and the WYMM parameter $\sigma$ that facilitates conditional exact solvability. That is, \[\Omega_0\geq0 \Rightarrow \tau_{max}=6\tilde\alpha^2+\sigma^2 \Rightarrow \ell_{max}=-\frac{1}{2}+\sqrt{\frac{1}{4}+6\tilde\alpha^2+\sigma^2},\] is the maximum allowed angular momentum quantum number. Hence, the values of the GM parameter $\tilde\alpha$ and the WYMM parameter $\sigma$ dictate the maximum allowed values for $\ell$. We should observe that for $\sigma=0$ and $0<\tilde\alpha<1$ allow $\ell=0$, while for $\sqrt{1/3}\leq \tilde\alpha<1$ allow $\ell=0,1$.

In Figure \ref{fig:2}, we show the energy for the  KG-oscillator in EiBI gravity and a WYMM. In Figures \ref{fig:1}(a), \ref{fig:1}(b), and \ref{fig:1}(c) we plot the energies against the Eddington parameter $\tilde\kappa$, where in \ref{fig:2}(a) we use $0<\tilde\alpha <1$ that allows $\ell=0$ and in \ref{fig:2}(b) we use $0.75<\tilde\alpha<1$ that allows $\ell=1$ for $\sigma=0$ (without WYMM). In \ref{fig:2}(c) we fix $\tilde\alpha=0.9$ and use different values for the WYMM parameter $\sigma$. In \ref{fig:2}(d) we fix $\sigma=3.5$ and $\tilde\kappa=0.5$ so that $\ell=0,1,2,3$ values are allowed. We may observe the exact symmetry of the energies of KG particles and antiparticles about $E=0$. We also observe that increasing the WYMM strength parameter $\sigma$ increases the energies for particles and antiparticles.

\section{Concluding remarks}\label{sec:5}

\setlength{\parindent}{0pt}

In this work, we investigated the nonrelativistic Schr\"odinger and relativistic KG oscillators in the spacetime of a GM within the framework of EiBI gravity. In the relativistic sector, we also incorporated the interaction with a WYMM. By introducing the non-minimal oscillator coupling and extracting an appropriate exponential prefactor, the radial equations were transformed into the confluent Heun form. The controlled termination of the Heun series then yielded conditionally exact solutions. When the series truncates at the first nontrivial order, the radial eigenfunctions reduce to closed form polynomials and the truncation conditions impose explicit algebraic relations among the oscillator frequency, the EiBI parameter, the monopole deficit parameter and the allowed orbital angular momenta.

\setlength{\parindent}{0pt}

The mathematical and physical structure of the resulting spectra is particularly transparent. For the Schr\"odinger oscillator, the termination condition produces the discrete energy levels \(E_{n}=2\,\omega\,\tilde{\alpha}^{2}\left(n+\frac{5}{2}\right),\) along with a nontrivial frequency constraint that depends on the combination \(\tau={\ell(\ell+1)}/{\tilde{\alpha}^{2}}\). This leads to a parametric correlation for $n=0$, \(\omega_{0}=\frac{1}{8\tilde{\kappa}}\left[10-5\tau+\sqrt{\tau^{2}-12\tau+100}\right],\)which fixes the admissible values of the oscillator frequency and simultaneously enforces an upper bound on the orbital quantum number $\ell$. This restriction, determined jointly by the EiBI parameter and the monopole deficit, is a characteristic manifestation of conditional exact solvability in curved backgrounds. Figures \ref{fig:1}(a)-\ref{fig:1}(f) show the resulting spectral behavior: the energies increase with the monopole parameter $\tilde{\alpha}$, decrease with the EiBI parameter $\tilde{\kappa}$, and follow precisely the analytically predicted limits. Figure~\ref{fig:1} summarizes the behavior of the lowest, $n=0$, state energy $E_{0,\ell}$ obtained from the analytic expression in Eq.~(\ref{2.15}) for a wide range of $\tilde{\kappa}$ and $\tilde{\alpha}$. Panels (a)--(c) show that for fixed $\tilde{\alpha}$ the quantity $E_{0,\ell}$ increases with $\tilde{\kappa}$, indicating that the nonlinear contribution of EiBI strengthens the effective potential. The enhancement becomes more pronounced as $\tilde{\alpha}$ grows and the case $\ell=1$ displays the additional expected upward shift caused by the centrifugal term. Panels (d)--(f) demonstrate that for fixed $\tilde{\kappa}$ the energy rises monotonically with $\tilde{\alpha}$, with the growth rate controlled by the magnitude of $\tilde{\kappa}$. Thus, Figure~\ref{fig:1} shows that the nonlinear structure of the EiBI GM spacetime produces a consistent upward modification of the Schr\"odinger oscillator energy throughout the explored parameter domain.  

\setlength{\parindent}{0pt}

In the KG system, the same mechanism leads to the symmetric energy branches \(E_{0}=\pm\sqrt{10\,\tilde{\alpha}\,\Omega_{0}+m_{0}^{2}}\), where the frequency $\Omega_{0}$ depends sensitively on the GM charge $\sigma$ and the geometric parameters of the background. The WYMM modifies the effective angular momentum and produces a well defined upper limit \(\ell_{\max}=-\frac{1}{2}+\sqrt{\frac{1}{4}+6\tilde{\alpha}^{2}+\sigma^{2}}\), which determines the number of admissible bound states. Figures \ref{fig:2}(a)-\ref{fig:2}(d) clearly reveal these effects: the energy levels of the particles and antiparticles remain perfectly symmetric about zero, increase with the monopole strength, and adjust predictably as $\tilde{\kappa}$ and $\tilde{\alpha}$ vary.
In Figure \ref{fig:2}, the four panels collectively show how the EiBI GM gravitational deformation $(\tilde{\alpha},\tilde{\kappa})$, the nonlinear coupling $\sigma$, and the orbital quantum number $\ell$ govern the relativistic energy branches of the KG oscillator in the WYMM framework. Panels (a) and (b) reveal that for $\sigma=0$ the energy increases monotonically with the parameter $\tilde{\kappa}$, while larger values of $\tilde{\alpha}$ shift both the positive and negative branches upward in magnitude. This shows that EiBI deformation enhances the effective confining strength of the system. The transition from $\ell=0$ to $\ell=1$ in panel (b) produces a noticeable elevation of the spectrum at the same $\tilde{\alpha}$, reflecting the growing influence of the angular momentum barrier in curved spacetime. In panel~(c), fixing $\tilde{\alpha}=0.9$ and increasing the nonlinear parameter $\sigma$ significantly amplifies the energy magnitude, indicating that $\sigma$ controls an additional geometric rigidity induced by the WYMM extension. The curves separate more strongly with $\sigma$, showing that nonlinear effects become dominant for higher orbital modes. Panel~(d) exhibits the dependence on $\tilde{\alpha}$ for multiple $\ell$ values, where higher orbital states systematically experience larger energy shifts. In all subfigures, the positive and negative branches remain perfectly symmetric, reflecting the relativistic particle-antiparticle nature of the KG oscillator. The results show that EiBI gravity and the WYMM nonlinear contribution introduce substantial modifications to the oscillator spectrum, providing a "tunable" gravitational mechanism that reshapes relativistic bound-state energies.

\setlength{\parindent}{0pt}

\textcolor{red}{Before concluding, a comparison with previous works on oscillators in global monopole and EiBI backgrounds is unavoidable.} \textcolor{red}{In so doing, we recall the $(n+1)th$-order three-term recursion relation (\ref{2.12})
\begin{equation}
\begin{split}
A_{n+3}\, \left( n+3\right)\left( n+\frac{5}{2}\right) &=A_{n+2}\left[ \left(
n+2\right) \left( n+P+1\right)-\mu \right] \\ 
&+A_{n+1}\left[ \mu+\nu-\frac{\tilde \kappa\omega}{2} (n+1)\right]=0;\\
&\,\, n=0,1,2,\cdots.
\end{split} \label{5.1}
\end{equation}
(which is a common result associated with the truncation of the power-series solution (\ref{2.10}) to a polynomial of degree $(n+1)$ for the confluent Heun equation). The truncation procedure may follow different approaches:
\begin{enumerate}
    \item One may assume that $A_{n+3}=0, A_{n+2}=0$ and $A_{n+1}\neq0$ which implies that \[ \mu+\nu-\frac{\tilde \kappa\omega}{2} (n+1)\Rightarrow  \mu=-\nu+\frac{\tilde \kappa\omega}{2} (n+1)\geq0,\] to consequently yield \[\delta=-\alpha\left[\tilde n+\frac{1}{2}\left(\beta+\gamma+2\right)\right],\]where $\tilde n=n+1\geq1$ is a positive integer, together with the condition that $\eta$ is a root of $A_{n+2}\equiv A_{n+2}(\eta)=0$, as in Ronveaux \cite{rf34}, which is generally difficult to determine.
    \item Alternatively, many authors (e.g., \cite{rf21,rf23,rf32}) consider only the lowest state, $n=0\Rightarrow \tilde n=1$, and use only the $n{th}$ order recursion relation \[A_{n+1}\left[ \left(n+1\right) \left( n+P\right)-\mu \right] 
+A_{n}\left[ \mu+\nu+\alpha n\right]=0,\]along with \[A_1=-\frac{\mu}{\beta+1} A_0;\quad A_0=1,\] to determine the eigenvalues, while ignoring the first condition in (\ref{5.1}). This procedure leads to incorrect results as reported by Ahmed and Bouzenada \cite{Ref32} for the Schr\"odinger oscillator, so that their reported result in their Eqs. (19)-(24) and (62)-(67) are valid only for positive signatures (including their reported figures). 
\item Recently, Mustafa et al. \cite{rf24,rf25,rf26} have suggested the use of the $nth$ order recursion relation, associated with  (\ref{2.11}):
\begin{equation}
\begin{split}
A_{n+2}\, \left( n+2\right)\left( n+\frac{3}{2}\right) &=A_{n+1}\left[ \left(
n+1\right) \left( n+P\right)-\mu \right] \\ 
&+A_{n}\left[ \mu+\nu-\frac{\tilde \kappa\omega}{2} n\right]=0.
\end{split} \label{5.2}
\end{equation}
and have used the truncation conditions $A_{n+2}=0, A_{n+1}\neq0$ and $A_n\neq0$, where the coefficients of $A_{n+1}$ and $A_n$ identically vanish. In this case, one obtains \(\mu+\nu=\frac{\tilde \kappa\omega}{2} n\) and \(\left(
n+1\right) \left( n+P\right)=\mu\). This approach yields conditional exact solvability not only for $n=0$ but also for all $n$ values.
\item In the current methodical proposal, we were motivated by the unusual positive and negative signatures that appeared in the spectroscopic structure of the non-relativistic Schr\"odinger oscillators reported in \cite{Ref32}. We have shown that not only the negative signatures are invalid and incorrect in \cite{Ref32}, but also the oscillator potential, $V(\rho)=\Omega^2\rho^2=\frac{1}{4}\omega^2\rho^2\,;\,\rho^2=r^2+\tilde \kappa$, they have used, is not the correct potential induced by EiBI gravity spacetime but should rather be corrected to read\[V(\rho)=\frac{1}{4}\omega^2\rho^2+\frac{\omega\rho^2}{\rho^2-\tilde \kappa},\]upon the change of variable $\rho^2=r^2+\tilde \kappa$ in our (\ref{2.6}). The second term constitutes a significant EiBI gravitational field contribution that has been dismissed by the authors in \cite{Ref32}.
\end{enumerate}}

\setlength{\parindent}{0pt}

Several directions naturally follow from this work. Extending the approach to spinor fields or to other curved backgrounds where Heun structures arise may uncover additional analytically tractable sectors. The method presented here provides a coherent and analytically controlled framework for obtaining exact and conditionally exact solutions of oscillator systems in EiBI GM geometries and in the presence of magnetic monopole fields. The results enrich the set of solvable quantum models in curved and topologically nontrivial spacetimes and lay a solid foundation for further theoretical and numerical investigations.

\end{document}